\documentclass{caosp308}

\usepackage{graphicx}
\usepackage{adjustbox}
\usepackage{subcaption}
\usepackage{amsmath}
\usepackage{amssymb}
\usepackage{multirow}
\usepackage{array}
\usepackage{verbatim}
\usepackage[section]{placeins}
\usepackage{natbib}
\newcommand{\TESS}{\textit{TESS}}

\newcommand{\pdcsap}{\textsc{PDCSAP}}
\articleNo{123}
\pubyear{2020}
\volume{35}
\volnumber{3}
\firstpage{1}
\received{September 15, 2020}
\accepted{November 27, 2020}

\def\BibTeX{{\rm B\kern-.05em{\sc i\kern-.025em b}\kern-.08em
             T\kern-.1667em\lower.7ex\hbox{E}\kern-.125emX}}

\begin{document}

\hauthor{J.\,Korth et al.}
\title{Consequences of parameterization choice on eclipsing binary light curve solutions}

\author{
       J.\,Korth\inst{1}\orcid{0000-0002-0076-6239}  
      \and 
         A.\,Moharana\inst{2}
      \and 
          M.\,Pe\v{s}ta\inst{3}
      \and
         D.R.\,Czavalinga\inst{4}    
      \and 
         K.E.\,Conroy\inst{5}\orcid{0000-0002-5442-8550}
       }

\institute{ 
           Rheinisches Institut f\"ur Umweltforschung, Abteilung Planetenforschung an der Universit\"at zu K\"oln, Universit\"at zu K\"oln, Aachenerstraße 209, 50931 K\"oln, \email{judithkorth@gmail.com}
         \and 
           Nicolaus Copernicus Astronomical Center, Polish Academy of Sciences, ul. Rabia\'{n}ska 8, 87-100 Toru\'{n}, Poland, \email{ayushm@ncac.torun.pl}
         \and 
           Institute of Theoretical Physics, Charles University, V Hole\v{s}ovi\v{c}k\'{a}ch 2,
           \mbox{180 00 Praha 8, Czech Republic},
           \email{pestamilan@gmail.com}
         \and
           Institute of Physics, University of Szeged, 6720 Szeged, Hungary, \email{czdonat@titan.physx.u-szeged.hu}
           \and
           Department of Astrophysics and Planetary Science, Villanova University, 800 East Lancaster Avenue, Villanova, PA 19085, USA
          }

\date{September 15, 2020}

\maketitle

\begin{abstract}
Eclipsing Binaries (EBs) are known to be the source of most accurate stellar parameters, which are important for testing theories of stellar evolution. With improved quality and quantity of observations using space telescopes like {\it TESS}, there is an urgent need for accuracy in modeling to obtain precise parameters. We use the soon to be released \texttt{PHOEBE 2.3} EB modeling package to test the robustness and accuracy of parameters and their dependency on choice of parameters for optimization.  
\end{abstract}

\section{Introduction} \label{intro}
It is well known that eclipsing binaries (EBs) provide highly accurate observations of stellar parameters, which is important for testing theories of star evolution. Thanks to the increasingly more precise photometry of the past and recent space missions (e.g. \textit{Kepler} \citep{borucki2016} and the Transiting Exoplanet Survey Satellite \citep[\TESS;][]{tess2014}), it is now possible to observe and study EBs in details never seen before (e.g. reflection from the companion, lensing, or Doppler beaming). To understand the observations, one needs to employ models capable of making predictions with sufficient precision, so that it is possible to compare the observations with predicted theoretical values. A popular example of such software is \texttt{PHOEBE} \citep{phoebeprsa, phoebehov, phoebejones}, a robust Python package for modeling of EB systems. The latest release provides users with control over a large number of orbital and physical parameters, which allows them to generate synthetic light curves and radial velocities of the binary system. One can also take advantage of various built-in or imported solvers (e.g. \texttt{emcee}) and optimizers (e.g. Nelder-Mead) to solve the inverse problem---for a comprehensive introduction to the inverse problem using \texttt{PHOEBE} see \citet{kylephoebe2.3}.

For the purpose of this paper, we use the soon to be released version 2.3 of \texttt{PHOEBE} to try and reproduce the results from the article by \citet{maxted2020}, which examines a number of various methods to accurately estimate the masses and radii for the stars in the binary system AI Phoenicis (AI~Phe). This system, which contains two well-separated, sharp-lined stars of comparable luminosity, was first identified as an EB by \citet{strohm}. It is an excellent target for model testing as it is relatively bright ($V = 8.6$\,mag), has a long orbital period ($P\approx24.59$\,days), and does not show any distinct spots nor flares associated with increased magnetic activity of the components \citep[e.g.,][]{kkent2016, maxted2020}.

To compare the results by \citet{maxted2020}, we first carried out a number of runs with varying underlying physical models, free parameters, and their initial values (see Section~\ref{dataset}). This further motivated us to design a controlled experiment in order to systematically analyze the effect of parameterization choice on the final light curve resulting from the model (Section \ref{contrexp}). Finally, in Section~\ref{discussion}, we compare our results from the controlled experiment with the \citet{maxted2020} values, and we discuss our findings regarding the precision of the employed model.

\section{Observations and Modeling Set-up} \label{dataset}

The photometric data of AI Phe used in the subsequent analyses were obtained under the \TESS{} Guest Investigator Program (G011130, P.I. Maxted; G011083, P.I. Helminiak; G011154, P.I. Prša) during Sector 2 of the \TESS{} mission observed in the 2-min cadence mode (TIC 102069549). The Sector was observed for 27 days from $2458354.113259$ BJD to $2458381.517643$ BJD (covering both the primary and secondary eclipse), and the data were reduced by the \TESS{} data processing pipeline developed by the Science Processing Operations Center \citep[SPOC;][]{jenkins2016}. In our analyses, we used the Pre-search Data Conditioning Simple Aperture Photometry (\pdcsap) light curve, which was additionally detrended by fitting a chain of $5^{\rm th}$ order Legendre polynomials \citep[][Section 2.6]{maxted2020}.

To get a sense of the effect of parameterization on the resulting values, we independently solved the inverse problem for AI Phe by using a separate model with its own set of free parameters and approximation of physical phenomena (e.g. limb-darkening law, reflection, etc.). Following the approach from \citet{maxted2020}, we initialized the parameters of the models with their estimates from \citet{kkent2016}, which are summarized in Table \ref{kentdata2}.

\begin{table}[h!] 
\caption{A list of the parameters of the binary system AI Phe that were adopted from \citet{kkent2016}.}
\label{kentdata2}
\centering
\begin{tabular}{ll}
\hline
\hline
Parameters  & Values\\
\hline
$P$ (days) & 24.592483  \\
$q$ & 1.0417   \\
$e$ & 0.1821  \\
$\omega(^{o}$ )& 110.73 \\
$i(^{o}$ ) & 88.502 \\
$M_{1}$ ((M$_{\odot}$) & 1.1973 \\
$M_{2}$ ((M$_{\odot}$)& 1.2473 \\
$R_{1}$ ((R$_{\odot}$)) & 1.835 \\
$R_{2}$ ((R$_{\odot}$))&  2.912 \\
$T_{1}$ (K) & 6310 \\
$T_{2}$ (K)&  5237.3 \\
\hline
\hline
\end{tabular}
\end{table}

The initialized free parameters were used as input to the Nelder-Mead algorithm \citep{nmead} in order to refine the estimates. These estimates then served as a starting point for initial distributions (either Gaussian or uniform) of the free parameters entering the Markov Chain Monte Carlo (MCMC) algorithm implemented in the \texttt{emcee} solver \citep{emcee}, which we used to obtain posterior distributions of the relevant parameters. Furthermore, we used the following software: \texttt{Python} \citep{python}, and the \texttt{Python} libraries \texttt{Matplotlib} \citep{matplotlib} and \texttt{numpy} \citep{numpy}. 

Unfortunately, the individual runs did not yield satisfactory results as the obtained values showed a wide spread. Due to the different choices of parameterization and the large numbers of parameters, it was not possible to associate the observed variation with a specific parameter or a set of parameters. Therefore, we decided to design a controlled experiment, in which we defined a ``nominal" run and then we examined the effect of altering the parameters one at a time. For more information see the following section.

\section{Controlled Experiment} \label{contrexp}
The ``nominal" run (which we shall denote ``Run A") served as a benchmark for all the other runs (``B" through ``K"), which we systematically varied from Run A in a controlled fashion---that is, for each run, we altered one aspect of the ``nominal" set-up and kept the rest unchanged. For the definitions of Runs B--K, see Table \ref{runstab}.

\begin{table}[h!] 
\caption{A list of the individual runs and their differences from the ``nominal" run.}
\label{runstab}
\centering
\resizebox{\textwidth}{!}{\begin{tabular}{ll}
\hline
\hline
Run  & Description\\
\hline
A & The ``nominal" run  \\
B & Logarithmic limb darkening law\\
C & Sample/interpolate in phase-space \\
D & Marginalization over albedos \\
E & Marginalization over gravity darkening parameters\\
F
& Marginalization over gravity darkening parameters from \\ &  \citet{gravdcoeff} \\
G
& Marginalization over noise nuissance parameter\\
H
& Marginalization over parameters $q$ and $a$ using radial velocities \\ & posteriors from \citet{rvdata} on $q$ and $a\sin{i}$\\
I & Meshes on binary surfaces to estimate $L_\mathrm{pb}$\\
J & \TESS{} light curve without detrending (PDCSAP)\\
K
&  Masking the out-of-eclipse points \\
\hline
\hline
\end{tabular}}
\end{table}

\begin{figure}[!htb]
\begin{subfigure}{0.5\textwidth}
\leftline{
\includegraphics[width=\linewidth]{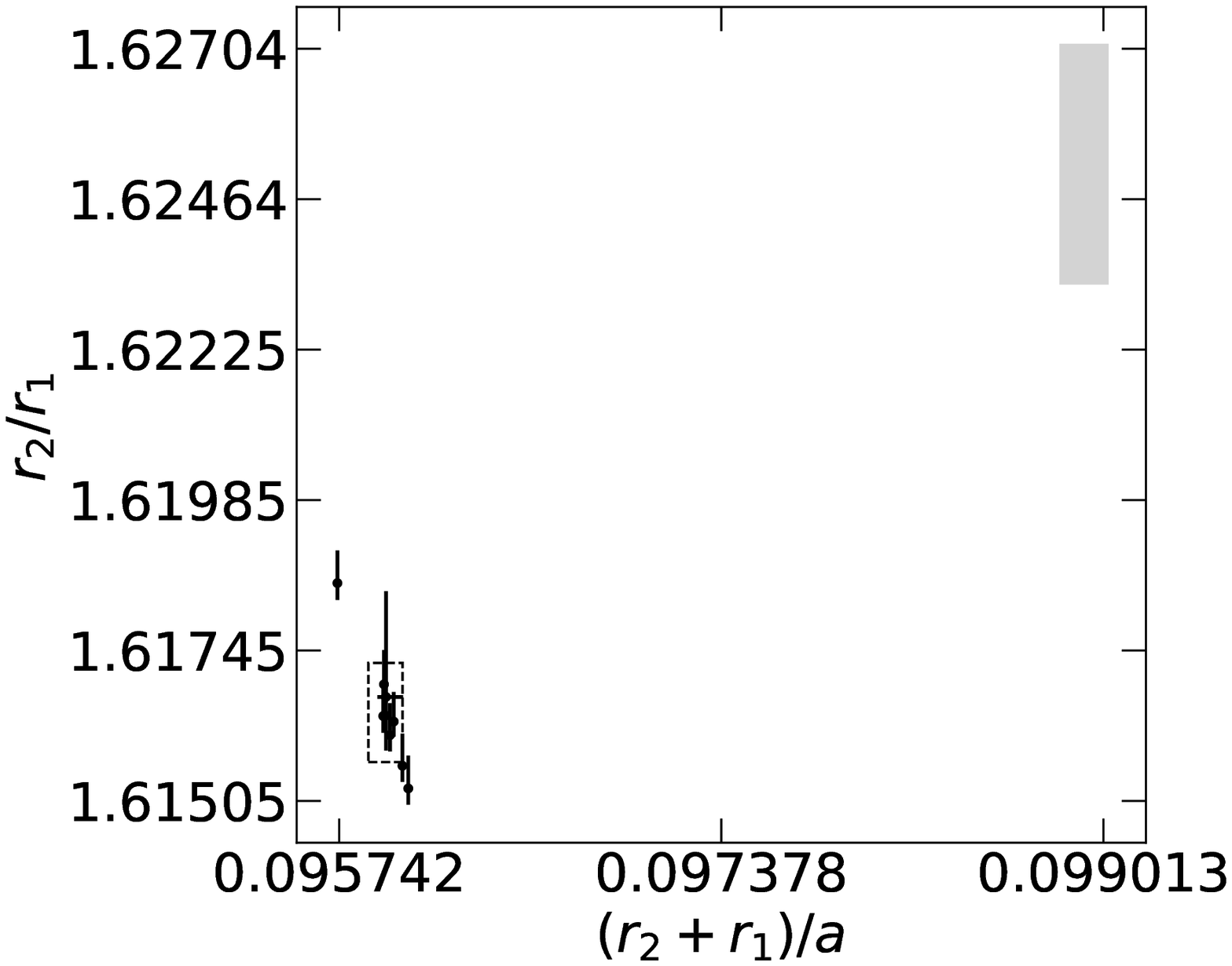}
}
\label{fig:a}
\end{subfigure}
\begin{subfigure}{0.5\textwidth}
\leftline{
\includegraphics[width=\linewidth]{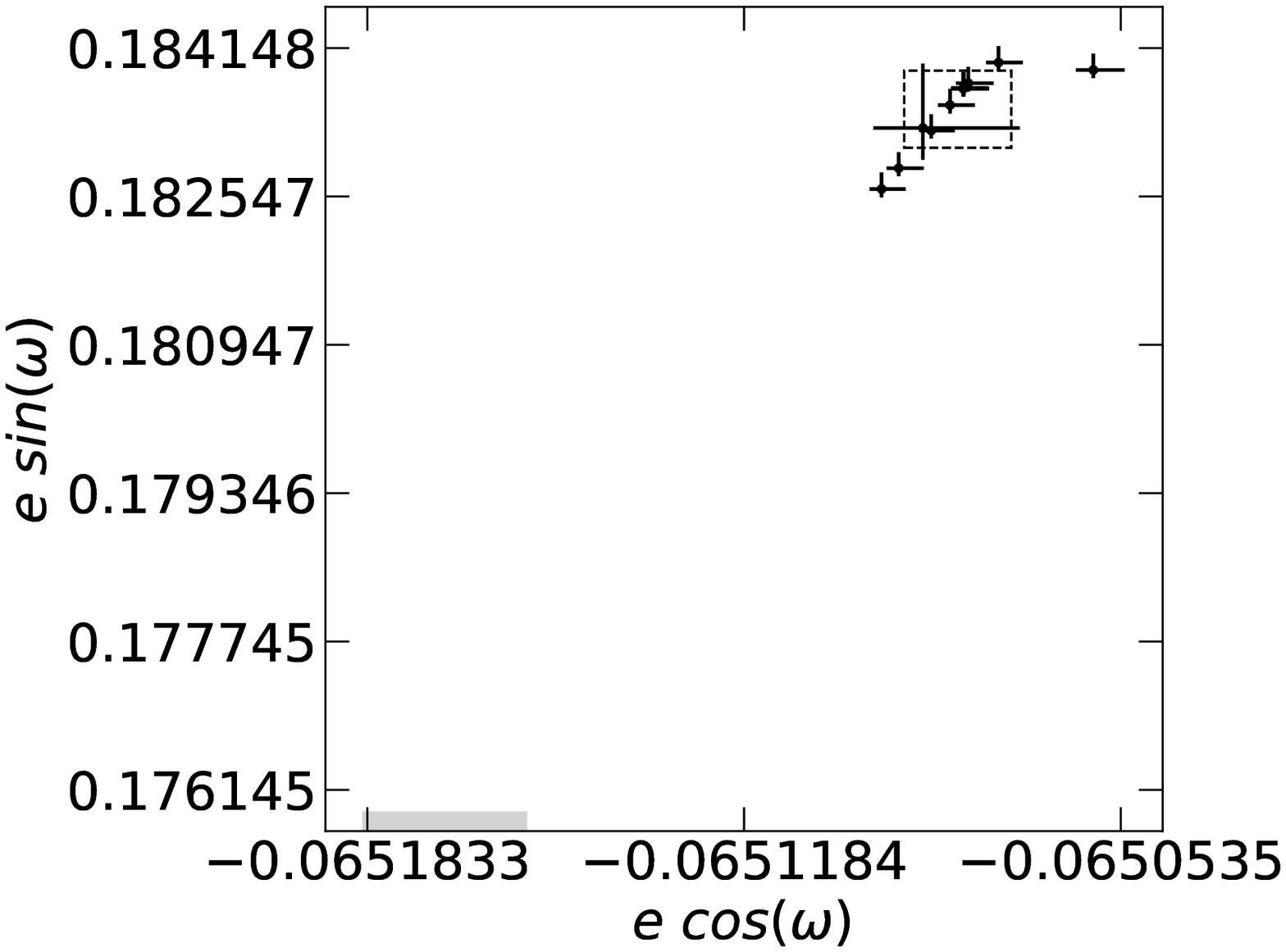}
}
\label{fig:b}
\end{subfigure}

\vspace{0.2cm}
\begin{subfigure}{0.5\textwidth}
\leftline{
\includegraphics[width=\linewidth]{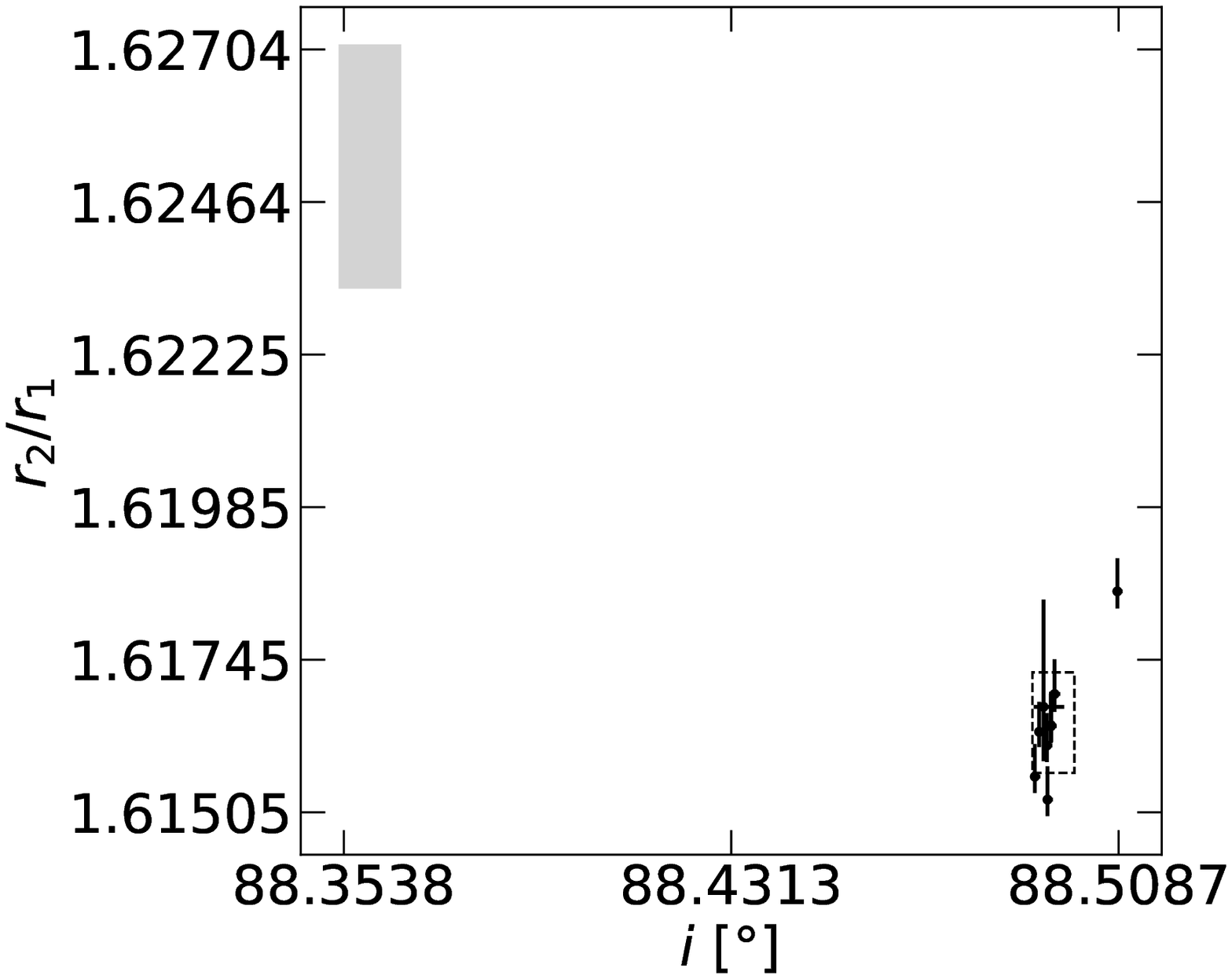}
}
\label{fig:c}
\end{subfigure}
\begin{subfigure}{0.5\textwidth}
\leftline{
\includegraphics[width=\linewidth]{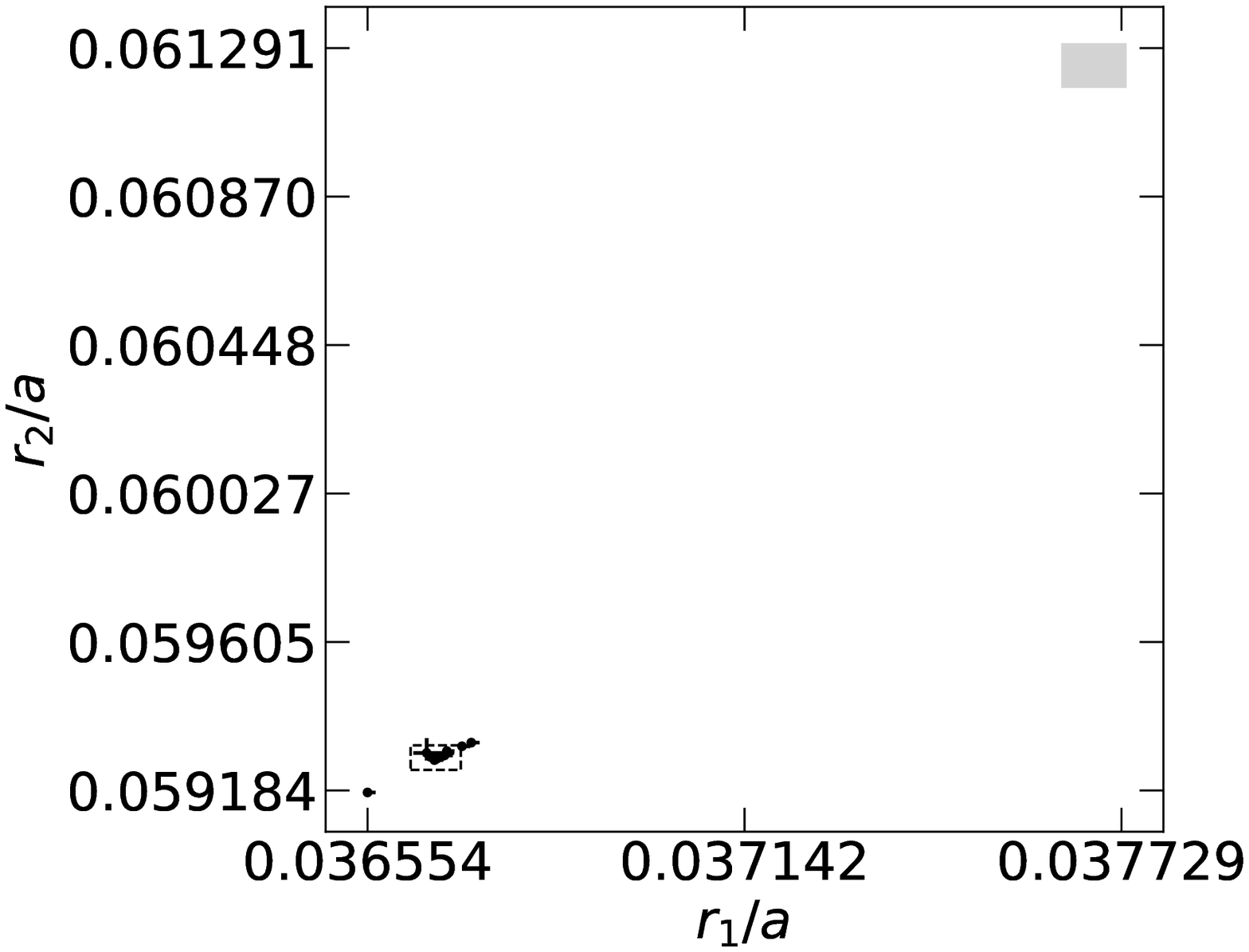}
}
\label{fig:d}
\end{subfigure}

\vspace{0.2cm}
\begin{subfigure}{0.5\textwidth}
\leftline{
\includegraphics[width=\linewidth]{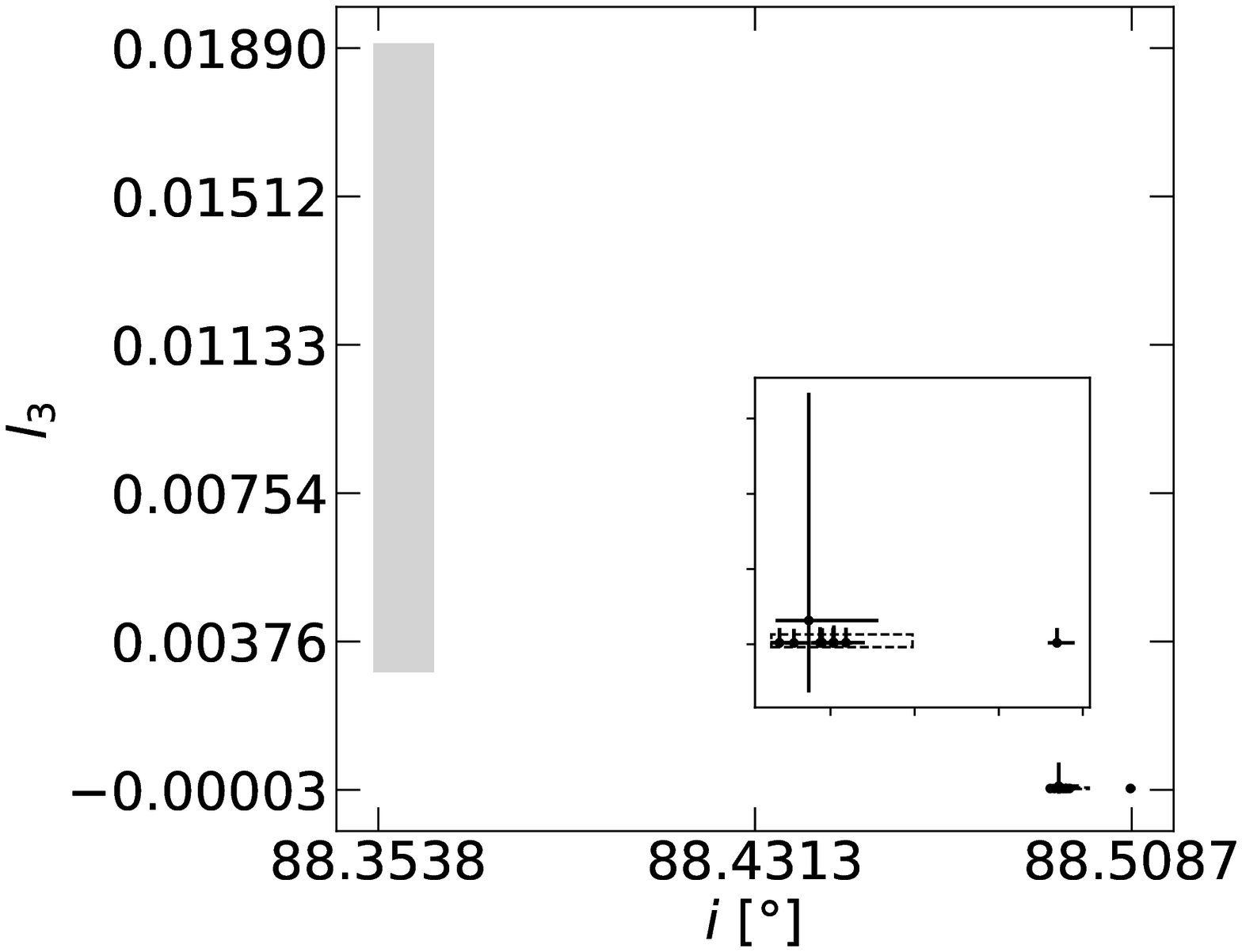}
}
\label{fig:e}
\end{subfigure}
\begin{subfigure}{0.5\textwidth}
\leftline{
\includegraphics[width=\linewidth]{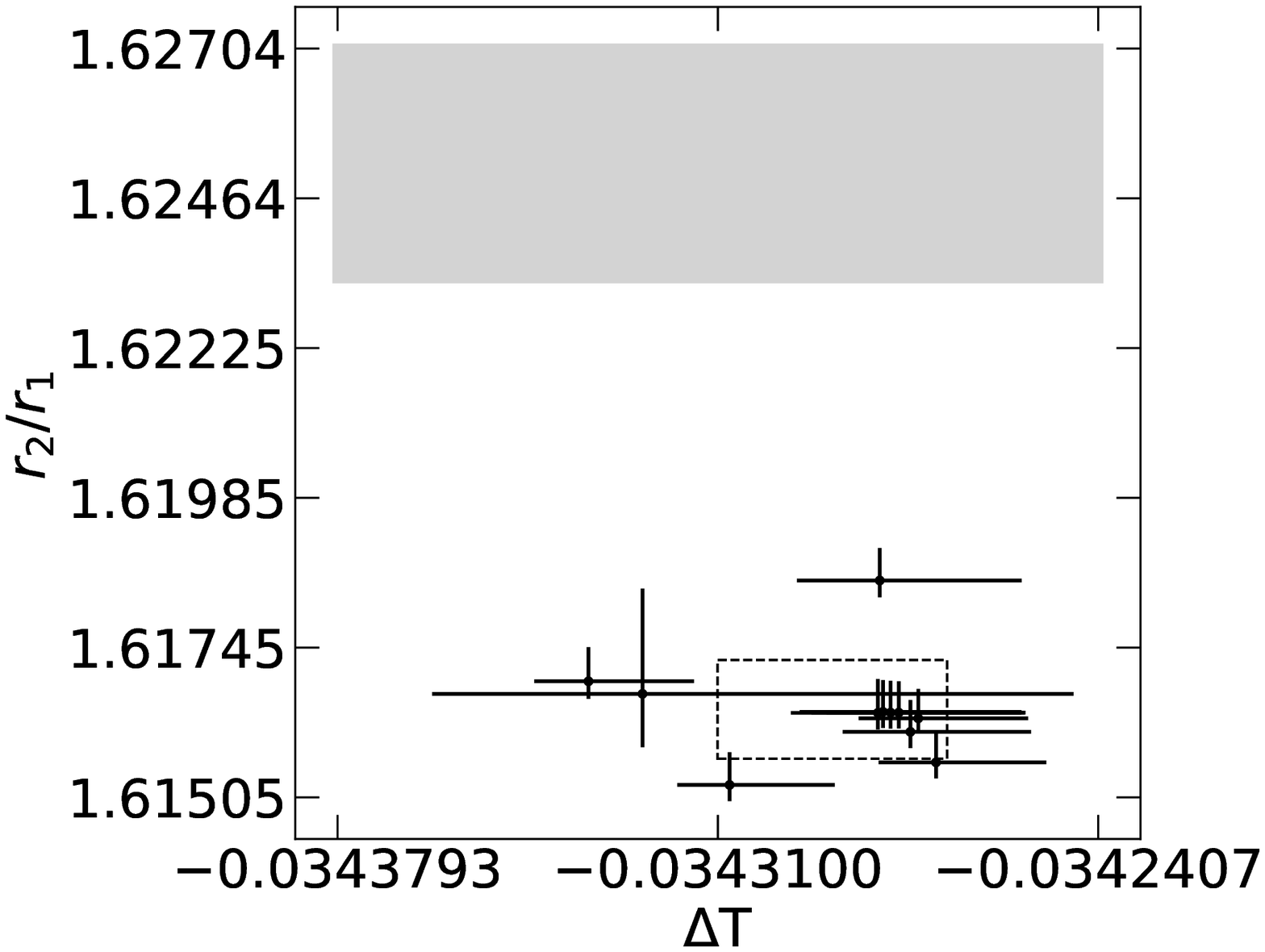}
}
\label{fig:f}
\end{subfigure}
\caption{A comparison between the parameters resulting from Runs A--K and those obtained by \citet{maxted2020}. The filled rectangle represents the 1-$\sigma$ spread around the average of the runs from the Maxted et al. paper (see Table \ref{tab3}), while the dotted box shows the same for our controlled runs. The inner box in the $l_{3}$ vs $i$ plot represents a zoomed-in view on the  parameter distribution.}
\label{fig:comparison}
\end{figure}

Run A uses the binary star model \textsc{ellc} \citep[for more information see][]{ellc} with the quadratic limb-darkening law in the ``lookup" mode (automatic querying of coefficients from tables based on mean stellar values), and uses the Stefan-Boltzmann approximation in the determination of the passband luminosity, $L_\mathrm{pb}$, which is needed to scale the fluxes and estimate the surface-brightness ratio. Similar to our initial test runs, we initialized the parameters with values from \citet{kkent2016}, and then used the Nelder-Mead algorithm to refine the estimates. After that, we used the \texttt{emcee} solver to sample over the radii, $R_1$ and $R_2$, (for the primary and secondary component), the eccentricity, $e$, along with the argument of pericenter, $\omega_{0}$, (parameterized as $e \sin{\omega_{0}}$ and $e \cos{\omega_{0}}$), the time of the primary eclipse, $T_{0}$, the third light, $l_{3}$, $L_\mathrm{pb}$, the ratio of the effective temperature of the secondary and primary component, $T_\mathrm{secondary}/T_\mathrm{primary}$, and the orbital inclination, $i$, to get an estimate for their uncertainties. We present the obtained results in Fig. \ref{fig:comparison}.

\section{Discussion of Results} \label{discussion}
Analysis of Runs A--K shows a rather limited spread of the obtained values, with the individual parameters lying within each other's uncertainties (see Fig.~\ref{fig:comparison}). Seeing that the parameters have a fairly small effect on the final results, this implies that the high variation of our initial runs was most likely not caused by any specific parameter but rather by a combined effect of various parameterization choices.  That said, it is still clear that the choices made in Table \ref{runstab} do influence the final results.

As for the runs presented in \citet{maxted2020}, each of them employs a distinct combination of the underlying physical model, optimization method and parameterization (see Table \ref{tab3}). In principle, to properly compare our results with those obtained by \citet{maxted2020}, the initial set-ups should also be compared so that the effect of initial configuration choice can be distinguished from other effects. Although we compare our results with all the runs listed in Table \ref{tab3} (see Fig.  \ref{fig:comparison}), we shall inspect in detail only the initial set-ups for Runs A and S by Maxted, as they utilize the same physical binary model (\texttt{ellc}) and optimization method (\texttt{emcee}) as our runs. Our runs use a wrapper for mapping of \texttt{PHOEBE} parameterization onto \texttt{ellc}), and thus they minimize the effect of choice (there is still the freedom of parameterization) and can serve as a ``benchmark" for our results. Moreover, Runs A and S by Maxted use essentially the same initial set-ups, therefore it suffices to examine only the former (which additionally corrects for instrumental systematic variations). To avoid confusion, we shall prefix the runs by \citet{maxted2020} with ``M-" (e.g. Run A by Maxted becomes Run M-A) in the rest of the section.

For the sake of simplicity, we shall compare the parameterization of Run M-A only with our ``nominal" Run A, which is assumed to be representative of the other runs (for their definitions, see Table \ref{runstab}). In Table \ref{tab4}, we compare the free parameters entering the \texttt{emcee} algorithm in the two runs. Apart from working with mostly disjunct (but correlated) sets of free parameters, the runs also differ in the limb-darkening law (power-2 for Run M-A and quadratic for Run A) and the treatment of the stellar masses. Run A adopts the values 1.1973\,M$_{\odot}$ and 1.0417 from \citet{kkent2016} for the mass of the primary component and the ratio of the masses of the secondary and primary component, respectively. In addition, the stellar masses are held constant at the Kirkby-Kent values as they have minimal effect on the light curve. In contrast, Run M-A uses the \texttt{emcee} posterior distributions of the free parameters together with observed radial velocities of the binary system to get an estimate for the masses. Finally, both runs hold the semi-major axis constant, with Run M-A assuming the value of 47.868 R$_{\odot}$ and Run A keeping it equal to 47.941 R$_{\odot}$.

\begin{table}[h!] 
\caption{An overview of the various runs analysed in the paper by \citet{maxted2020}. The table was adopted from page 6 of the mentioned paper.}
\label{tab3}
\centering
\resizebox{\textwidth}{!}{\begin{tabular}{llllll<{\raggedleft}m{4cm}}
\hline
\hline
Run & Investigator & Model & Optimization & Limb-darkening & Detrending & Notes\\
\hline
A & Maxted & \texttt{ellc} & \texttt{emcee} & power-2 & \texttt{celerite} & \\
B & Helminiak & \texttt{JKTEPOB} & L-M & quadratic & sine+poly & Monte Carlo error estimated\\
C & Torres & \texttt{EB} & \texttt{emcee} & quadratic & spline & Quadratic l.d. coeffs. fixed\\
D & " & " & " & " & " & \\
E & Graczyk & \texttt{WD2007} & L-M & logarithmic & -- & Fixed l.d. coefficients\\
F & Johnston & \texttt{PHOEBE 1.0} & \texttt{emcee} & square-root & -- & \\
G & Prša & \texttt{PHOEBE 2.1} & MCMC & grid & legendre & \\
H & Orosz & \texttt{ELC} & DE-MCMC & logarithmic & polynomial & \\
I & Orosz & " & " & square-root & " & \\
J & Orosz & " & " & quadratic & " & \\
K & Southworth & \texttt{JKTEBOP} & L-M & quadratic & polynomial & \\
L & Southworth & \texttt{JKTEBOP} & L-M & cubic & polynomial & \\
S & Maxted & \texttt{ellc} & \texttt{emcee} & power-2 & \texttt{celerite} & Same as Run A with SAP light curve\\
\hline
\hline
\end{tabular}}
\end{table}

\begin{table}[h!] 
\caption{A list of the free parameters entering the \texttt{emcee} algorithm in Runs M-A and A. The ``Nelder-Mead" column marks the parameters which were optimized before running \texttt{emcee}, in order to speed-up the convergence of the posterior distributions.} \label{tab4}
\centering
\resizebox{\textwidth}{!}{\begin{tabular}{l<{\raggedleft}m{7cm}|ccc}
\hline
\hline
\multirow{2}*{Parameter} & \multirow{2}*{Description} & Run M-A & \multicolumn{2}{c}{Run A}\\
&& \texttt{emcee} & Nelder-Mead & \texttt{emcee}\\
\hline
$e \cos{\omega}$ &  &  & \checkmark & \checkmark\\
$e \sin{\omega}$ &  &  & \checkmark & \checkmark\\
$f$ & Flux scaling factor & \checkmark &  &\\
$f_c$ & $\sqrt{e}\cos{\omega}$ & \checkmark &  &\\
$f_s$ & $\sqrt{e}\sin{\omega}$ & \checkmark &  &\\
$h_\text{1,F}, h_\text{2,F}$ & Parameters of the power-2 limb darkening law for star 1 & \checkmark &  & \\
$h_\text{1,K}, h_\text{2,K}$ & Parameters of the power-2 limb darkening law for star 2 & \checkmark &  & \\
$i$ & Orbital inclination & \checkmark & \checkmark & \checkmark\\
$k$ & Ratio of the radii & \checkmark &  &\\
$l_3$ & Third light & \checkmark &  & \checkmark\\
$L_\mathrm{pb}$ & Passband luminosity &  &  & \checkmark\\
$r_{sum}$ & Sum of the fractional radii & \checkmark &  &\\
$R_1$ & Radius of the primary star &  & \checkmark & \checkmark\\
$R_2$ & Radius of the secondary star &  & \checkmark & \checkmark\\
$\sigma_f$ & Standard error per observation & \checkmark &  &\\
$S_{\text{T}}$ & Surface brightness ratio averaged over the stellar disks in the TESS band & \checkmark &  &\\
$T_\text{0}$ & Time of primary eclipse & \checkmark & \checkmark & \checkmark\\
$T_\text{secondary}/T_\text{primary}$ & Ratio of the effective temperatures &  & \checkmark & \checkmark\\
\hline
\hline
\end{tabular}}
\end{table}

\begin{figure}[h!]
    \centering
    \includegraphics[width=0.7\linewidth]{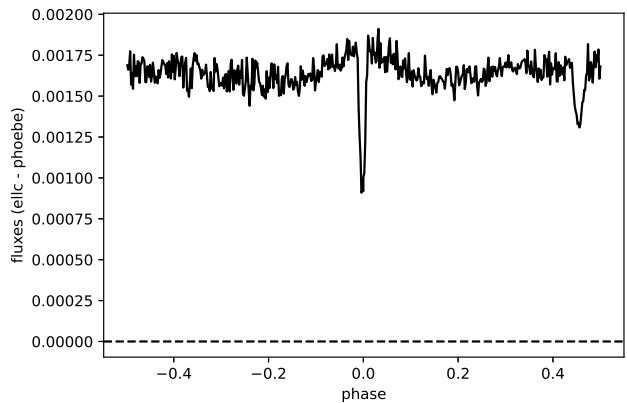}
    \caption{Residuals between the \texttt{ellc} and \texttt{PHOEBE} forward models for the nominal case.  This shows the influence of the approximations in mapping between the parameterization of the different codes and likely explains the offset between the results from our controlled sample and those in \citet{maxted2020}.  }
    \label{fig:model_resids}
\end{figure}

Coming back to the general case, we expected our results to agree with those obtained by \citet{maxted2020} within the reported uncertainties despite the differences in the initial set-ups of the runs. However, Fig.~\ref{fig:comparison} shows that this is not the case as none of the results from Runs A–K lies in the 1-sigma spread of the \citet{maxted2020} values. As to the reason behind this discrepancy, multiple explanations present themselves. First, due to time and computational constraints, we stopped our \texttt{emcee} runs after appearing flat (converged to a single value) for about 1500 iterations compared to $\sim$10000 iterations for the runs in \citet{maxted2020}. Thus, there is a slight possibility that the runs might yet ``jump" and converge to some other values. All of our runs, however, were treated consistently with each other, and still exhibit the influence of a number of decisions in the fitting process on the final results. Next, the mapping between the \texttt{PHOEBE} and \texttt{ellc} parameterizations includes several assumptions and approximations, which leads to residuals between the two forward models (see Fig.~\ref{fig:model_resids}). This suggests that the offset between the two sets of results might be caused by not employing the native \texttt{PHOEBE} backend.

\section{Conclusions}
In this work, we tried to fit the parameters of AI Phoenicis from the \TESS{} light curve and to reproduce the values estimated by \citet{maxted2020}.
First, we independently modeled the light curves with individual sets of free parameters and their approximations, e.g., difference between initialized parameters of binary masses, radius, effective temperatures of the stars, or model of limb darkening, reflections etc. Since our independent models did not lead to the same results for the system parameters of AI Phe, we designed a controlled experiment to systematically analyze the effect of the parameterization. 

The parameters obtained from different runs were expected to be distributed in a parameter space as found in \cite{maxted2020} but we find our results to be quite different in comparison to the accuracy that has been found before. We suspect the discrepancy may be reconciled by running \texttt{emcee} for substantially longer to allow further convergence and to switch to the native \texttt{PHOEBE} forward model to avoid the assumptions and approximations in the translation between parameterizations.  Our results do, however, show the importance of several different choices in the fitting process on the final parameter values and their uncertainties.

\acknowledgements

The work presented here was part of the GATE Summer School, hosted (virtually) by Masaryk University Brno and sponsored by ERASMUS+ under grant number 2017-1-CZ01-KA203-035562. The authors would like to thank all involved and Marek Skarka for his organization. A.M. would like to acknowledge the support provided by the Polish National Science Center (NCN) through the grant number 2017/27/B/ST9/02727.

\bibliography{bibliography}

\end{document}